\documentclass{PoS}

\title{A deep survey of the low-surface-brightness radio sky}

\ShortTitle{A low-surface-brightness radio survey}

\author{\speaker{Ravi Subrahmanyan}\\
        Raman Research Institute, C.V. Raman Avenue, Sadashivanagar, Bangalore
        560080, India\\
        ATNF-CSIRO, PO Box 76, Epping NSW 1710, Australia\\
        E-mail: \email{rsubrahm@rri.res.in}}

\author{R. D. Ekers\\
        ATNF-CSIRO, PO Box 76, Epping NSW 1710, Australia\\
        E-mail: \email{Ron.Ekers@csiro.au}}

\author{Lakshmi Saripalli\\
       Raman Research Institute, C.V. Raman Avenue, Sadashivanagar, Bangalore
        560080, India\\
        ATNF-CSIRO, PO Box 76, Epping NSW 1710, Australia\\
        E-mail: \email{lsaripal@rri.res.in}}

\author{E. M. Sadler\\
       School of Physics, University of Sydney, NSW 2006, Australia\\
       E-mail: \email{ems@physics.usyd.edu.au}}

\abstract{We have made a radio survey---the Australia Telescope Low Brightness
  Survey (ATLBS)---of 8.4 square degrees sky area, using the
Australia Telescope Compact Array in the 20-cm band, in an observing mode
designed to provide wide-field images with exceptional sensitivity in surface
brightness, and thereby explore a new parameter space in radio source
populations.  The goals of this survey are to quantify the distribution in
angular sizes, particularly at weak surface brightness levels: this has
implications for the confusion in deep surveys with the SKA.  The survey is
expected to lead
to a census of the radio emission associated with low-power radio galaxies at
redshifts 1--3, without any missing extended emission, and hence a study of
the cosmic evolution of low-power radio galaxies to higher redshift and a
comprehensive study of the AGN feedback during the intense black hole growth
phase during this redshift range. }

\FullConference{From planets to dark energy: the modern radio universe\\
		 October 1-5 2007\\
		 University of Manchester, Manchester, UK}

\begin{document}

\section{Introduction}

Radio continuum surveys are characterised by an observing frequency and a
limiting flux density at which the survey is deemed to be more or less
complete.  An important aspect which is often overlooked is the angular
resolution (or, more precisely, the spatial frequency coverage in the case of
surveys made using interferometers) of the survey telescope and the variation
in the flux sensitivity with the angular size of the source.  Single-dish
radio telescopes have large beams and,
therefore, their sensitivity is limited by confusion.  Interferometer surveys
with good flux sensitivity are `blind' to extended source components and may
miss sources with low surface brightness.  As a consequence, population
studies of extragalactic radio sources are incomplete in their understanding
of the low surface brightness sources.

Table~1 lists existing surveys that have been made with
particularly good surface brightness sensitivity.  To compare surveys we have
assumed a frequency dependence $S_{\nu} \propto \nu^{-1}$ of the flux density
$S_{\nu}$ with frequency $\nu$, typical for extended components of
extragalactic radio sources.  Additionally, we have assumed that the extended
source sizes exceed the beam (an optimistic assumption for the low
resolution surveys). The surface brightness detection limit equivalent at the
frequency of 1.4~GHz, which are listed in the last column of the table,
correspond to 5-$\sigma$ limits in all cases except the 6C survey,   
which is confused, and for which an optimistic limit to the 151-MHz flux
density for reliable detection of extended sources is 1~Jy (Saunders et
al. 1987).  

\begin{table}[b]
\begin{tabular}{lccccc} 
\hline 
Survey & Telescope & Frequency & 1-$\sigma$ & Beam FWHM &  
Surface brightness \\ 
 & & & RMS noise & & (mJy/arcmin$^2$) \\ 
\hline 
\hline 
6C &  Cambridge & 151~MHz & 20~mJy & 4.2$^{\prime} \times 4.2^{\prime}$cosec$\delta$ & 6  \\ 
7C & Cambridge & 151~MHz & 15~mJy & 70$^{\prime\prime} \times 70^{\prime\prime}$cosec$\delta$ & 6  \\ 
WENSS & WSRT & 326~MHz & 3.6~mJy & 54$^{\prime\prime} \times 54^{\prime\prime}$cosec$\delta$ & 5 \\ 
B2 & Bologna & 408~MHz & 50~mJy & 5$^{\prime} \times 4^{\prime}$sec(ZA) & 3.6 \\ 
B3 & Bologna & 408~MHz & 20~mJy & 2.6$^{\prime} \times 4.8^{\prime}$sec(ZA)  & 2.3 \\ 
SUMSS & MOST & 843~MHz & 1~mJy & 43$^{\prime\prime} \times 43^{\prime\prime}$cosec$\delta$ &  6 \\ 
NVSS &  
VLA & 1.4~GHz & 0.45~mJy & 45$^{\prime\prime}$ & 4  \\ 
PMN & Parkes & 4850~MHz & 5~mJy & 4.2$^{\prime}$ & 7 \\ 
\hline 
\end{tabular}
\caption{The equivalent surface brightness detection limit of surveys computed
  at a frequency of 1.4~GHz. ZA denotes zenith angle, and FWHM is the full
  width at half maximum of the telescope beam.}
\label{table_1}
\end{table}  

As may be seen from Table~1, none of these surveys could  
have detected sources with surface brightnesses below a few  
mJy~arcmin$^{-2}$.  Deep small-area radio surveys like the HDF-S and  
the Phoenix deep fields recently observed with the ATCA reach
surface-brightness   
levels of 0.3--0.4~mJy~arcmin$^{-2}$, as do similar surveys made with  
the VLA with arcsec resolution and $\mu$Jy flux sensitivity.   
These surveys can detect extended emission from high-redshift objects, but  
because the total area surveyed is small, they cover too small a volume  
to give a fair sample of the low-surface-brightness radio source population.
 
\section{The ATLBS survey}

The new radio survey covers 8.4 square degrees of sky with a surface
brightness 
sensitivity good enough to detect, at 5-$\sigma$, arcmin-scale extended
sources with flux density down to 0.4~mJy.  The survey has been given the
acronym ATLBS - the Australia Telescope Low Brightness Survey. In Fig.~1 we
compare the 
5-$\sigma$ limiting surface brightness sensitivity equivalent at 1.4~GHz for
the different surveys versus angular size; ATLBS is about
a factor 5 better in surface brightness sensitivity compared to any
previous survey with comparable resolution.    

\begin{figure}[b]
\centering
\includegraphics[angle=-90,width=10.5 cm]{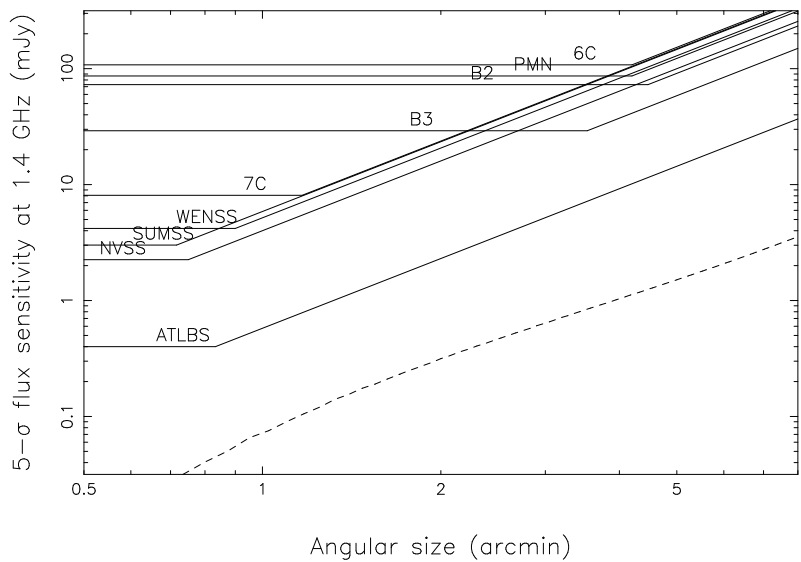}
\caption{Limiting surface-brightness sensitivity  
(equivalent at 1.4~GHz) for various surveys versus source angular size.
The survey resolution and  
flux sensitivity for earlier surveys are as listed in Table~1.   
The dashed line has been drawn at 5 times the R.M.S. confusion noise  
for a fully-filled visibility sampling, and 
shows that the ATLBS survey is well above the confusion limit.}
\label{fig_1}
\end{figure}

High fidelity surveys for extended sources with low surface brightness 
requires good spatial frequency coverage.  Holes in the u,v-coverage 
effectively reduce the number of independent synthesized beam areas 
within the primary beam and, consequently, confusion owing to  
discrete sources limits the image dynamic range and quality.  

The ATLBS 
observations used the 750A, 750B, 750C and 750D array configurations of the
E-W Australia Telescope Compact Array (ATCA).  
These together provide $4 \times 10 = 40$ baselines and 
because the ATCA antennas are 22-m in diameter, the 40 spacings 
provide a nearly complete coverage over the 0--750~m range.   
At the 750-m baseline, 
Earth rotation moves the visibility point through 22~m in about 7~min. 
Therefore, our observing strategy covered 19 fields in every 12-hr observing  
session: each field was observed for 20~sec so that the 19 fields 
were re-visited once every 7~mins and each of the fields have, 
consequently, complete azimuthal coverage along the u,v-tracks. 
The blank time between successive pointings was about 3~sec 
and this resulted in a 15\% loss of integration time. The 19 fields 
tile the sky in a hexagonal pattern, with pointings 
spaced 28.6~arcmin on the sky so as to cover the sky with uniform 
sensitivity. 

In this special mode, every field was observed at a centre frequency of 1388~MHz
using a pair of 128-MHz bands with dual polarization. 
The synthesized beam FWHM is 50~arcsec for naturally weighted data. 
The observations are in full 
polarization mode resulting in full polarization images. The 
images have R.M.S. noise of $80~\mu$Jy~beam$^{-1}$, and the 5-$\sigma$ 
detection limit is 0.4~mJy~beam$^{-1}$.   

It may be noted here that we have not truly mosaic imaged
the fields: what we  
have done is a `cut-and-paste' mosaic of a significant sky area with 
exceptional (and uniform) surface brightness sensitivity;  
however, the survey is
relatively insensitive to sources with angular extents near the 
telescope primary beam FWHM size of 35$^{\prime}$. 
The four 750-m arrays will yield images with about 1$^{\prime}$ resolution;  
however, the visibility coverage is complete so 
good quality images with lower 
resolution may be synthesized from the data.  The good u,v coverage
enables  
imaging all structures up to 10$^{\prime}$ in size with the 5-$\sigma$ 
surface brightness  
sensitivity of 0.4~mJy~beam$^{-1}$. 
 
The ATCA observations in the 750-m arrays include baselines to
a `6-km antenna': the long baselines to this antenna provides 
the high resolution that identifies compact components 
so that the presence of any cores and/or hotspots of extragalactic  
double sources would be distinguished 
from the extended source components. The long spacings 
are crucial in distinguishing the cases where closely spaced 
point sources appear as slightly resolved `extended' sources in 
the images with 1$^{\prime}$ beams.    
 
We would like to emphasize the importance of the complete u,v-coverage 
that is possible with the ATCA in the mode described above.  
Good and complete visibility coverage is vital  
to avoid confusion from the stronger sources above the detection limit.
If the 
u,v-coverage were a factor of two less well filled, the effective synthesized
beam  
area that would respond to weak source confusion would double, and the 
R.M.S. confusion owing to the weak sources would double. The complete 
u,v-coverage that may be achieved with the fast position switching possible at 
the ATCA is crucial to confusion-free imaging at these 
low-surface-brightness levels. 
 
\section{Preliminary results}

Two regions---each a mosaic of 19 fields---have been observed centred at RA:
$00^{\rm h}~35^{\rm m}~00^{\rm s}$, DEC:
$-67^{\circ}~00^{\prime}~00^{\prime\prime}$ and RA: 
$00^{\rm h}~59^{\rm m}~17^{\rm s}$, DEC:
$-67^{\circ}~00^{\prime}~00^{\prime\prime}$ (J2000 
  epoch), covering a
total of 8.4~deg$^{2}$ sky area.  The final synthesized mosaic images,
made with beam FWHM of 50$^{\prime\prime}$, have no artefacts above 3-$\sigma$ of the
thermal noise: the dynamic range---defined as the ratio of the peak intensity
to the peak artefact anywhere in the image---exceeds 1000. A portion of the
survey is shown in Fig.~2.  Thus, we reliably detect and interpret contours at
5-$\sigma$ (0.4~mJy~beam$^{-1}$). Additionally, by constraining the
deconvolution algorithm to detect sources in the regions where the low surface
brightness survey identified sources, we have successfully used the baselines
to the `6-km antenna' to independently image the compact components in all of
the sources with a 5$^{\prime\prime}$ FWHM beam.

\begin{figure}[t]
\centering
\includegraphics[width=12 cm]{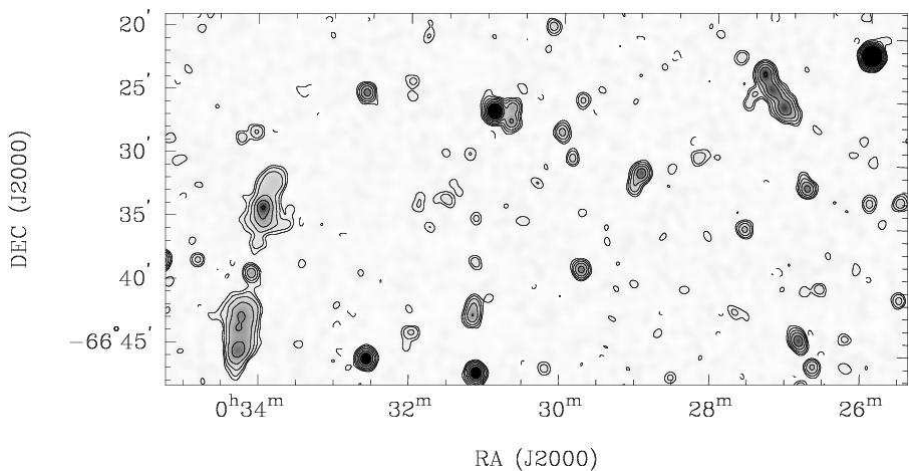}
\caption{Portion of the ATLBS survey.  The image was made with beam FWHM $52.4
  \times 47.4$~arcsec$^{2}$ at position angle 6$^{\circ}$. Contours at
  0.25~mJy~beam$^{-1} \times (-2$, $-1$, 1, 2, 4, 8, 16, 32, 64, 128).}
\label{fig_2}
\end{figure}

An immediate and obvious result of the survey, which demonstrated its quality,
was the discovery of a 1.9-Mpc giant radio galaxy J0034$-$6639 (seen in
Fig.~2). What is
remarkable is that the lobe surface brightness is lower than any other giant
radio galaxy known to date and has a value similar to that of the diffuse
radio halo in the Coma cluster; nevertheless, J0034$-$6639 is detected in
ATLBS at 40-$\sigma$.

We detect 1094 sources exceeding 0.4~mJy. At 50$^{\prime\prime}$ resolution, 10\% of the
sources have size exceeding 1.5 times the beam area, indicating that a
significant population of extended sources exist at mJy flux-density levels.
500 sources exceed 1~mJy: this sample represents a complete flux-limited
sample because it includes extended sources whose peak may be as low as
0.4~mJy. 
145 of these sources, or 30\% of the 1-mJy sample, have more than 50\% of the
total flux 
density in extended emission  that is detected in the low-surface-brightness
images with 50$^{\prime\prime}$ beam, but 
missing in the 5$^{\prime\prime}$ resolution images.

A few extended radio sources are coincident with nearby interacting and
star-forming systems, a few are halo-type systems identified with faint
clusters of galaxies.  Largely they are extragalactic radio sources associated
with AGN-type activity: radio cores are often detected that contain a large
fraction of the total flux density; however, a significant number have a large
fraction in extended emission.  Some extended sources have radio cores
detected coincident with optical counterparts in digital sky 
survey (DSS) images, some have radio
cores but no obvious optical IDs in DSS, some have optical IDs close to their
centroids but no core detected in the higher resolution images.

\section{Ongoing work}

A first paper presenting that ATLBS low-surface-brightness images and initial
results is under preparation.  Higher resolution radio observations with the
ATCA is underway to image the structures of all the extended mJy sources. We
are also mosaic imaging the fields in the K$_{\rm s}$ NIR band with the 3.9-m
Anglo-Australian Telescope, to improve identifications.

\acknowledgments

The ATCA is part of the Australia Telescope, funded by the Commonwealth of
Australia for operation as a National Facility managed by CSIRO.

\end{document}